\documentclass[pra,aps,10pt,superscriptaddress,twocolumn,floatfix]{revtex4-1}
\usepackage{graphicx,color}
\usepackage[nice]{nicefrac}							
\usepackage{amsmath,amssymb,bm}
\usepackage{ulem}

\usepackage[plainpages=false,pdfpagelabels,colorlinks=true,linkcolor=red,urlcolor=blue,citecolor=blue,pdftitle={},pdfauthor={},pdfdisplaydoctitle=true,pdfduplex=DuplexFlipLongEdge]{hyperref}

\definecolor{darkred}{rgb}{0.90,0.2,0.2}
\definecolor{darkgreen}{rgb}{0,0.60,.2}
\definecolor{darkblue}{rgb}{0.1,0.3,1}
\definecolor{grey}{cmyk}{0,0,0,0.25}
\definecolor{orange}{cmyk}{0,0.6,0.8,0}

\begin{document}

\title{Relaxation mechanisms in a disordered system with the Poisson level statistics}

\author{Janez Bon\v ca}
\affiliation{Department of Physics, Faculty of Mathematics and Physics, University of Ljubljana, SI-1000 Ljubljana, Slovenia}
\affiliation{Department of Theoretical Physics, J. Stefan Institute, SI-1000 Ljubljana, Slovenia}
\author{Marcin Mierzejewski}
\affiliation{Department of Theoretical Physics, Faculty of Fundamental Problems of Technology, Wroc\l aw University of Science and Technology, 50-370 Wroc\l aw, Poland}

\begin{abstract}
We discuss the interplay between many-body localization and spin-symmetry. To this end, we study the time evolution of several observables in the anisotropic $t-J$ model. Like the Hubbard chain, the studied model contains charge and spin degrees of freedom, yet it has smaller Hilbert space and thus allows for numerical studies of larger systems. We compare the field disorder that breaks the $\mathbb{Z}_2$ spin symmetry and a potential  disorder that preserves the latter symmetry. In the former case, sufficiently strong disorder leads to localization of all studied observables, at least for the studied system sizes. However, in the case of symmetry-preserving disorder, we observe that odd operators under the $\mathbb{Z}_2$ spin transformation relax towards the equilibrium value at relatively short time scales that grow only polynomially with the disorder strength. On the other hand, the dynamics of even operators and the level statistics within each symmetry sector are consistent with localization. 
Our results indicate that localization exists within each symmetry sector for symmetry preserving disorder. Odd operators' apparent relaxation is due to their time evolution between distinct symmetry sectors.

\end{abstract}

\maketitle
\section{Introduction}
The many--body localization (MBL)\cite{gornyi05,basko_aleiner_06,oganesyan_huse_07,AbaninRMP} phenomena has been most frequently studied in one--dimensional (1D) disordered systems with either  charge or spin degrees of freedom \cite{barisic_prelovsek_10,luitz15,luitz16,torres15,sirker14,pal10,bera15,Hauschild_2016,Devakul2015,bertrand_garciagarcia_16,Husex2017,Doggen2018}.
Even though    research in this field  mainly focused on a few simplest prototype model Hamiltonians for MBL such as the disordered  XXZ model, the type of the transition and even the existence  of the MBL phase in the thermodynamic limit are  still under intense consideration\cite{suntajs_bonca_20a,suntajs2020, sirker_2020, Luitz20, Vasseur20,Polkovnikov20,sieran2021a,vidmar2021,Huse21,sels2021,Sirker21}. One of the specific open problems is the effect of various  symmetries on the existence of MBL phase\cite{Vasseur2016,Chandran2014}. There are reports that non--Abelian symmetry precludes MBL \cite{Vasseur2016,proto2017,vasseur2021} while other investigations  claim that the non-Abelian symmetry  
is protected by  MBL\cite{friedman2017}. 

Shifting focus  to more complex prototype models that contain charge and spin degrees of freedom, such as the 1D  Hubbard model with potential disorder, the existence of the full MBL phase is even more elusive. In Ref.~\cite{prelovsek16,sroda19,kozarzewski2019,protopopov2018} authors investigate  the time evolution of spin and charge imbalance as well as transport properties  in the Hubbard model. Their results are consistent with localized/nonergodic  charge degrees of freedom while due to the preserved SU(2) symmetry the spin degrees of freedom remain  delocalized/ergodic up to extremely large values of potential disorder.  Similar conclusions were drawn based on the statistics of adjacent energy levels \cite{mondaini15} and by counting  the maximal number of local integrals of motion\cite{mierzejewski2018}. Subdifusive time evolution of charge particles was found   in the related   $t-J$ model\cite{bonca17} with potential disorder.

The effect of symmetry preserving disorder has been addressed already in non--interacting one-dimensional random hopping systems. In the case of systems with chiral or sublattice symmetry where particles can hop only between even- and odd- lattice sites, there is a  diverging mean density of states at zero--energy  \cite{dyson1953,cohen1976,reidinger1978},  which can lead to the delocalization transition \cite{brouwer1998,furusaki_NPB,furusaki2000,evers2008}.
  
The main goal of this work is to compare  the effect of spin-symmetry-preserving and symmetry-breaking disorders  on the dynamics of charge and spin degrees of freedom. In the case of  potential disorder, the behavior of specific  spin degrees of freedom
 is inconsistent with  the full MBL state due to the   $\mathbb{Z}_2$ spin-symmetry.  This observation is based on relatively fast relaxation of spin observables
 that are odd under the later  $\mathbb{Z}_2$  spin transformation, i.e.,
the only non--zero matrix elements of these operators connect two distinct  symmetry sectors. While this observation    seems to preclude MBL state,
  the  statistics of  adjacent energy  levels computed within each symmetry sector at large potential disorder approaches Poisson statistics.

We have organized the manuscript as follows: in the introduction, we present the model and the numerical method; we also discuss how the symmetry properties of the model depend on the type of the disorder. Next, we present the time evolution of the charge and spin imbalance and present a simplified model that explains the unusual relaxation of the spin imbalance in the presence of the potential disorder. We further present time evolutions of various charge and spin correlation functions, followed by the analysis of  charge and spin entanglement entropies. 
Based on the spectral level statistics, we discuss the apparent inconsistency between the relaxation of the spin imbalance and the Poisson level statistics, both observed under the influence of strong potential disorder. 

\section{Model and method.} 
We investigate the 
anisotropic $t-J$ model  on a one--dimensional ring with $L$--sites and $N_f=L/2$ fermions in the total $S^z=0$ subspace in the presence of either   a random external magnetic field $h_i \in [-2W_s,2W_s]$ or random potential $\epsilon_i \in [-W_c,W_c]$ 
\begin{eqnarray}
H &=& -t_0\sum_{i,\sigma}  c^{\dagger}_{i,\sigma} c_{i+1,\sigma} + c.c. +  J_z \sum_{i}S^z_iS^z_{i+1}- n_i  n_{i+1}/4  \nonumber \\ &+& {J_\perp \over 2}\sum_i  S^+_i S^-_{i+1} + S^-_i S^+_{i+1} + \sum_i h_iS^z_i + \epsilon_i  n_i.
\label{ham}
\end{eqnarray}
%
The fermion  operators,  $ c_{i,\sigma} $ and $ n_i = \sum_\sigma  c_{i,\sigma}^\dagger  c_{i,\sigma}$, as well as  the spin operators, $S^{\pm,z}_i$, 
act in the Hilbert space spanned locally by only three states, $|0\rangle_i$, $|\uparrow \rangle_i$, $|\downarrow \rangle_i$. Absence of doubly-occupied states, $|\uparrow \downarrow  \rangle_i$, in the $t$-$J$ model allows studying
charge and spin dynamics for much  larger systems than it would be possible for the Hubbard chain. This property is  the main motivation for the choice of Hamiltonian.
We perform multiple  time-evolutions based on the  Lanczos  technique,   each evolution starting from a different set of either random ${h_i}$  or  ${\epsilon_i}$.  The main goal of this work is to compare the time evolution of spin and charge degrees  of freedom under the influence of two different types of disorder. For this reason we choose,  following Ref.\cite{mondaini15}, the  initial state that  possesses  a charge--density--wave order as well as a staggered  spin orientation configuration $\vert \Psi_0\rangle$ defined as  
\begin{eqnarray}
\vert \Psi_0\rangle &=& \vert \uparrow 0 \downarrow 0 \uparrow 0 \downarrow 0 \dots \rangle,\\
\vert \bar\Psi_0\rangle &=& \hat{U} \vert \Psi_0\rangle = \vert \downarrow 0 \uparrow 0 \downarrow 0 \uparrow 0 \dots \rangle,
\end{eqnarray}
while $\vert \bar\Psi_0\rangle$ represents a state with a globally reversed spin projections, $S^z_i$ , via the unitary transformation  $\hat{U}=\prod_i (1-  n_i+S^+_i +S^-_{i})$.  In addition, we compute the level statistics of the  energy spectra.  We typically take $N_r=500-1500$ realisations of the disorder. We measure time in units of $[1/t_0]$ and set $t_0=J_z=1$, and $J_\perp=1.5$.  

There exists a significant  difference  between the two systems under the influence of  potential ($W_c\not =0, W_s=0$) and field disorder ($W_s\not =0, W_c=0$). Since we have set $J_z\not = J_\perp$, the $SU(2)$ symmetry is broken even at $W_{c(s)}=0$. In the case of the potential  disorder and for   $S^z_\mathrm{tot}=0$,
 the Hamiltonian remains invariant under the $\mathbb{Z}_2$ spin-transformation, $\hat{U}$,
which is closely related to the  global $\pi$-rotation around the $x$--axis  \cite{santos_2013}.
Since $\hat{U} \hat{U} =\hat{U}  \hat{U}^{\dagger} =1$,   each eigenstate consists of either symmetric ($\hat{U}  \vert \Psi_{S} \rangle=  \vert \Psi_{S} \rangle$)  or antisymmetric ($\hat{U}  \vert \Psi_{A} \rangle= - \vert \Psi_{A} \rangle$) combination of states
that differ by a global reversal of $S^z_i$, 
  $\vert \Psi_{S/A}\rangle = (\vert \Psi \rangle \pm \hat{U} \vert \Psi \rangle)/2$. The Hamiltonian thus  separates into two blocks with equal number of symmetric and antisymmetric functions. 

\section{Charge and spin imbalance}
We start by presenting time propagation of charge and spin imbalance as defined by the following operators
\begin{eqnarray}
\hat P_c &=& {1\over N_f}\sum_{i=1}^L (-1)^{i+1} n_i \\
\hat P_s &=& {2\over N_f}\sum_{i=1}^L (-1)^{\sum_{j=1}^{i-1}  n_j}S^z_i
\end{eqnarray}
\begin{figure}
\centering
\includegraphics[width=\columnwidth]{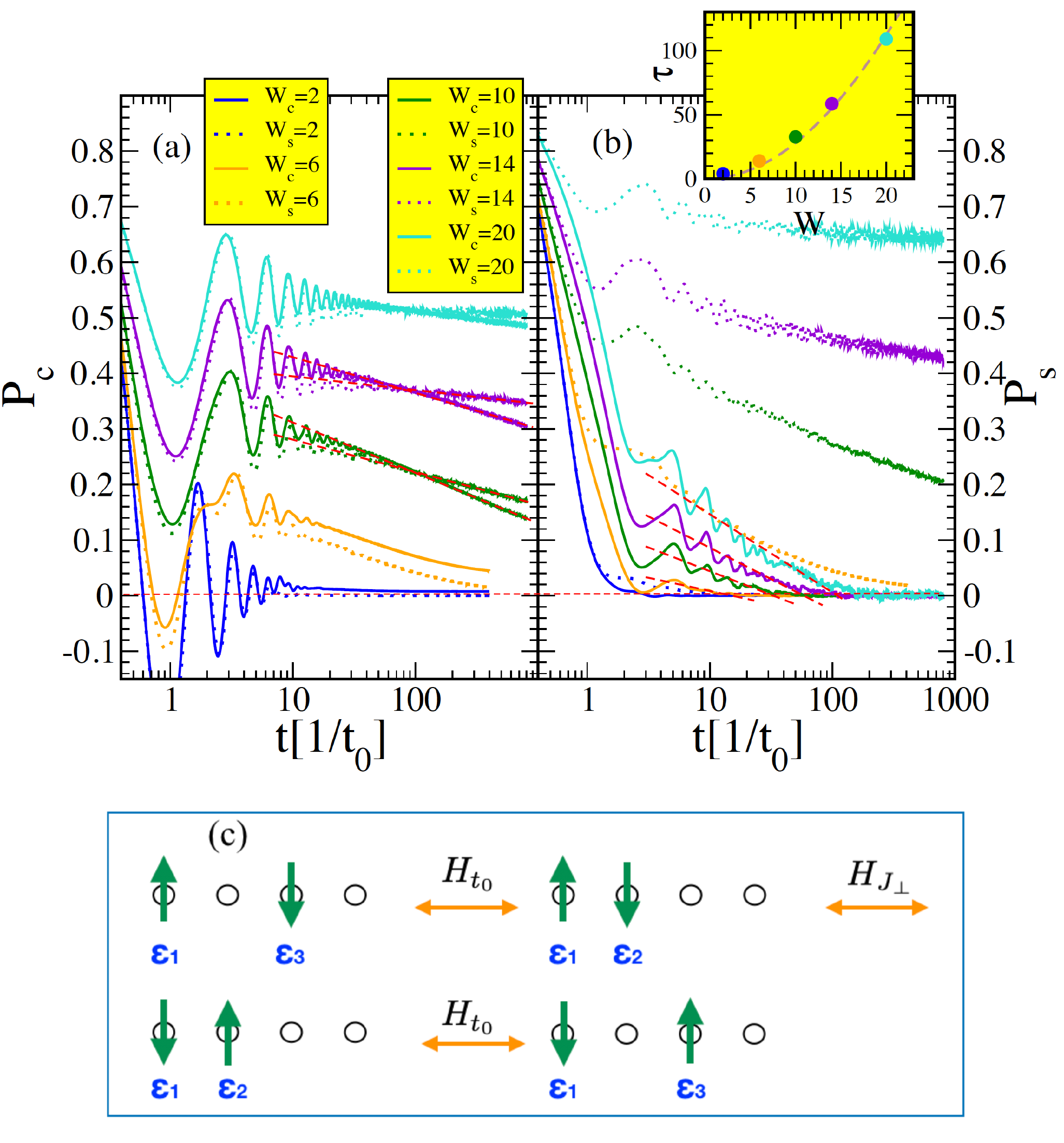}
\caption{$P_{c(s)}(t)=\langle\hat P_{c(s)}\rangle_t$ for a system with $L=16$. Note that $P_{s(c)}(t=0)=1$. Results using finite potential disorder $W_c$ (and $W_s=0$) are presented with full lines while those with field  disorder $W_s$ (and $W_c=0$) with dotted lines. Roughly $N_r\sim 500$ realizations have been used for each set of data. The inset in (b) represent the relaxation time $\tau$ defined through intercepts of $P_s$ with 0 obtained by fitting $P_s(t)=A\log(t)+B$ in the interval $t\gtrsim 4$ and $t\sim \tau_\mathrm{approx}$. Fits are displayed by thin dashed red lines. A brown  dashed line in the inset represents  a single--parameter  fit $\tau=0.28 W_c^2$. Similar fits as for $P_s(t)$  in a different time interval for $P_c(t)$ are also displayed in Fig.~\ref{Fig1}(a) using thin dashed red lines. Relaxation times for $W_{c(s)}=14$ based on presented fits are $\tau_c = 4.7\times 10^6$ and $\tau_s=7.7 \times 10^{16}$.  Small time step $\Delta t=0.01$ have been used in all presented results to obtain sufficient  numerical stability of time propagation. c) Processes that connect two  spin--reversed states with equal energy. Here,  $H_{t_0}$ and $H_\perp$ represent the first and the third term in Eq.~\ref{ham} for potential disorder.  }
\label{Fig1}
\end{figure} 
The  initial state is chosen such that $\hat P_{c(s)}\vert \Psi_0\rangle = \vert\Psi_0\rangle$.  Note also that $\hat P_s$ is odd under the $\mathbb{Z}_2$ spin-transformation, 
$\hat U \hat P_s \hat U =-  \hat P_s$,  thus it connects the symmetric with the antisymmetric sectors, while its matrix elements within each symmetry block vanish,  
$\langle \Psi_{S} \vert   \hat P_s \vert  \Phi_{S} \rangle=\langle \Psi_{A} \vert   \hat P_s \vert  \Phi_{A} \rangle=0$. 
Consequently, in the basis of the eigenstates of the Hamiltonian,   $\hat P_s$ has no diagonal matrix-elements.
In contrast $\hat P_{c}$  is even, $\hat U \hat P_c \hat U =  \hat P_c$  and  $\langle \Psi_{S} \vert   \hat P_c \vert  \Phi_{A} \rangle=0$.

We first show the charge imbalance, presented in Fig.~\ref{Fig1}(a), $P_c(t) = \langle \hat P_{c} \rangle_t$, where  $\langle \dots \rangle_t$  indicates multiple time evolutions from the initial $\vert \psi_0\rangle$, averaged over different  random realizations of either potential or magnetic field disorder. 
We observe  a similar time evolution for $t\lesssim 10$  under the influence of either potential  or field  disorder. At larger $t$ and for 
$W_{c(s)} \leq 6$, $P_c(t)$ relaxes towards zero faster  in the case of field  disorder. At larger  $W_{c(s)} \geq 10$
 $P_c(t)$ shows slow, logarithmic decrease in time. Providing that the time dependence would further follow a logarithmic form $A\log(t)+B$, as displayed with  thin dashed lines for $W_{c(s)}=10$ in Fig.~\ref{Fig1}(a), $P_c(t)$ would equilibrate under $W_c$ or $W_s$ to zero around $\tau_c\sim 27.000$ and $\tau_s\sim 460.000$, respectively. It is worth stressing that relaxation time becomes longer in the case of potential disorder  at larger values of $W_{c(s)}\geq 10$, moreover, relaxation might be prevented by the onset of the many body localization. 

   In the case of the spin imbalance $P_s(t)$, see Fig.~\ref{Fig1}(b), we find exceedingly  different time evolution when comparing  systems  with potential  or field  disorder. While $P_s(t)$ in the case of the field  disorder shows qualitatively similar behaviour as $P_c(t)$, the time evolution in the case of the potential disorder shows relaxation for all $W_s$ on a time scale accessible by our calculations. Moreover, the corresponding relaxation times $\tau$ show a quadratic $W_c$--dependence as shown in the inset of Fig.~\ref{Fig1}(b).  The latter quadratic dependence may be explained by recalling that for large $W_c, (W_s=0)$ a  redistribution of charge may be energetically very costly, however reversing the spin orientation does not change the energy.
Therefore the charge redistribution that is necessary for the spin dynamics is realized via virtual processes shown in  Fig.~\ref{Fig1}(c). In order to estimate the relevant energy scale (i.e. also the time-scale) one may study a toy model with four local states shown in  Fig.~\ref{Fig1}(c): two initial states $\vert \uparrow,0,\downarrow \rangle^{S/A}$ and two virtually-generated states $\vert \uparrow,\downarrow ,0\rangle^{S/A}$.  The corresponding eigenproblem  is up to a constant energy shift given by  $2 \times 2$ matrices
\begin{equation}
\left(
\begin{array}{cc}
0 & -t_0 \\ 
-t_0 & V^{S/A}
\end{array}
\right)
\left(
\begin{array}{c}
\vert \uparrow,0,\downarrow \rangle^{S/A}  \\ 
\vert \uparrow,\downarrow ,0\rangle^{S/A}\\ 
\end{array}
\right)
=E^{S/A} 
\left(
\begin{array}{c}
\vert \uparrow,0,\downarrow \rangle^{S/A}  \\ 
\vert \uparrow,\downarrow ,0\rangle^{S/A}\\ 
\end{array}
\right),
\label{exple}
\end{equation}
where the potentials $V^{S/A}=\epsilon_2-\epsilon_3-\frac{1}{2}(J_{z} \mp J_{\perp}) $. If the charge disorder is strong then the typical values of  $\vert V^{S/A} \vert \sim \vert \epsilon_2-\epsilon_3 \vert$ are large. As a consequence only two eigenstates have large projections on the initial states, $\vert \uparrow,0,\downarrow \rangle^{S/A}$.
Then, the dynamics of odd  operators is governed by the difference of corresponding eigenenergies in symmetric and antisymmetric sectors \mbox{ $\tau=|E^S-E^A|^{-1}\sim W_c^2/ t_0^2J_\perp $.}

\begin{figure}
\centering
\includegraphics[width=\columnwidth]{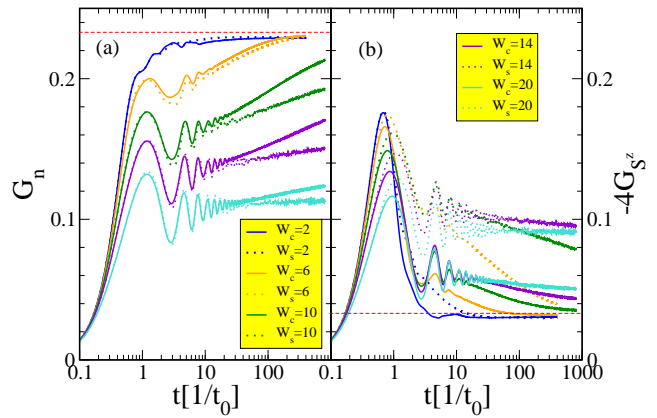}
\caption{  $G_n(t)$ in  (a) and $G_{S^z}(t)$ in (b) for different strengths of disorder. Full lines represent  potential disorder, $W_c$, while dotted lines magnetic field disorder, $W_s$. Thin horizontal red dashed lines represent infinite--$T$ limits of correlation functions, $G_n(T\to\infty)=7/30, G_{S^z}(T\to \infty)=-1/120$. Note, their respective values in the thermodynamic limit are $1/4$ and 0, respectively.  Parameters of the system are the same as in Fig.~\ref{Fig1}. }
\label{Fig2}
\end{figure} 
\section{Charge and spin correlation functions}
 We next explore the neighboring  density--density and spin--spin correlation function defined as:
\begin{eqnarray}
\hat G_n = {1\over L}\sum_i   n_i  n_{i+1}, \quad \quad G_n(t)=  \langle  \hat G_n \rangle_t,\\
\hat G_{S^z}={1\over L}\sum_i  S^z_i S^z_{i+1},\quad \quad G_{S^z}(t)=\langle   \hat G_{S^z}\rangle_t,
\end{eqnarray}
For the proper analysis of long--time behavior it is important to note that the  energy of the initial state after averaging over different random realizations, $E_\mathrm{ave}=\langle \psi_0\vert H\vert\psi_0\rangle_\mathrm{ave}=0$, is located near the middle of the energy spectrum, which in the microcanonical sense corresponds to infinite--temperature ($T\to \infty$). 
Based on the eigenstate thermalization hypothesis  \cite{srednicki_94, rigol_dunjko_08, deutsch_91, dalessio_kafri_16, eisert_friesdorf_15, deutsch_18, mori_ikeda_18} it is expected  that for small $W_{c(s)}$  and $t\to \infty$ $G_{n(S^z)}$ approach   their respective $T\to \infty$ limits, as indicated by dashed horizontal lines in Fig.~\ref{Fig2}.  It is indicative that charge and spin correlation functions for short times, $t\lesssim 1$ display qualitatively similar time dependence. Initially, only the hopping part  (the first term) of the Hamiltonian in Eq.~(\ref{ham}) is active since the exchange interaction can only act between  particles  on neighbouring sites. The change of either potential energy or field energy after hopping between neighboring lattice sites is in both cases comparable, which leads to a similar time--dependence for times comparable to inverse hopping time $1/t_0$.  For larger $W_{c(s)}\gtrsim 10$ and $t\gtrsim 50$,  $G_n$ shows logarithmic increase in time while $G_{S^z}$ shows logarithmic decrease. The difference is due to substantially different $T\to\infty$ limits. In contrast to $G_n$, $G_{S^z}$ shows  distinct dependence with regard to the  type of disorder. At large $W_c \geq10$,  $G_{S^z}$ approaches significantly  closer the ergodic $T\to \infty$ limit than in the case of $W_s \geq10$. Still, $G_{S^z}$ does not show relaxation as is the case of $P_s(t)$, shown in Fig.~\ref{Fig1}(b). The explanation for this seeming discrepancy can be found again in the symmetry argument. While $\hat P_s$ is odd under 
the  $\mathbb{Z}_2$ spin-transformation,  $\hat G_{S^z}$ is even, it has non-zero matrix elements within a fixed  symmetry sector, and the matrix elements  which are diagonal in the eigenstates of Hamiltonian may be non-zero as well.

\begin{figure}
\centering
\includegraphics[width=\columnwidth]{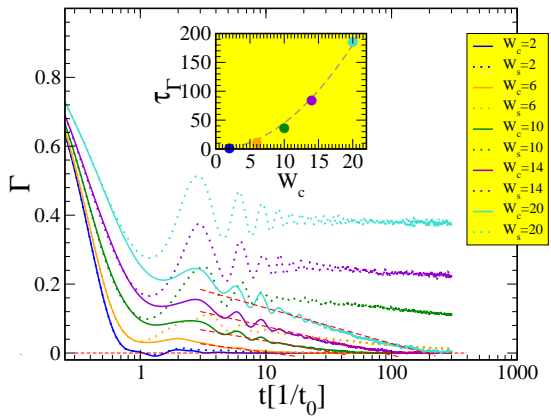}
\caption{  $\Gamma(t)=\langle\hat\Gamma\rangle_t$  for different strengths of disorder. Full lines represent  potential disorder $W_c$ while dotted lines magnetic field disorder $W_s$. Dashed red lines represent  fits of the form $\Gamma(t)=A \ln(t)+B$ that were used to extract relaxation times $\tau_\Gamma$, presented in the inset. Dashed brown line in the inset represents a single--parameter fit $\tau_\Gamma=0.45W_c^2$.  Parameters of the system are the same as in Fig.~\ref{Fig1}. }
\label{Fig3}
\end{figure}

To test this assumption,  we define a  three--site operator 
\begin{equation}
\hat \Gamma = {8\over N_f}\sum_{i=1,3,5,\dots,L-1}(-1)^{1+(i+1)/2} S^z_iS^z_{i+2}S^z_{i+4},
\end{equation}
which is also odd under the  $\mathbb{Z}_2$ spin-transformation, \mbox{$\hat U \hat \Gamma \hat U = - \hat \Gamma$}.
As seen in Fig.~\ref{Fig3}, $\Gamma(t) =\langle \hat \Gamma \rangle_t$ as well shows relaxation with the respective relaxation times $\tau_\Gamma$ scaling with $W_c^2$ just as in the case of $P_c(t)$, shown in Fig.~\ref{Fig1}(b). 

\section{Spin and charge entanglement entropy}

We now turn to  comparison  of the dynamics  of the entanglement
properties of charge and spin degrees of freedom. To this end we split the system into two halves. We then compute the charge and spin entanglement entropy\cite{lukin_rispoli_19,sirker_2020}, respectively:
\begin{eqnarray}
S_n &=& -\sum_{n=0}^{N_f} p_n \log(p_n)\\
S_{S^z}&=& -\sum_{S^z=-S^z_\mathrm{max}}^{S^z_\mathrm{max}} p_{S^z} \log(p_{S^z}),
\end{eqnarray}
where $S^z_\mathrm{max}=N_f/4$ and $p_n$ and $p_{S^z}$ represent probabilities that the subsystem  contains either   $n-$ fermions or  the $z-$component of the total spin equals $S^z$, respectively. 

\begin{figure}
\centering
\includegraphics[width=\columnwidth]{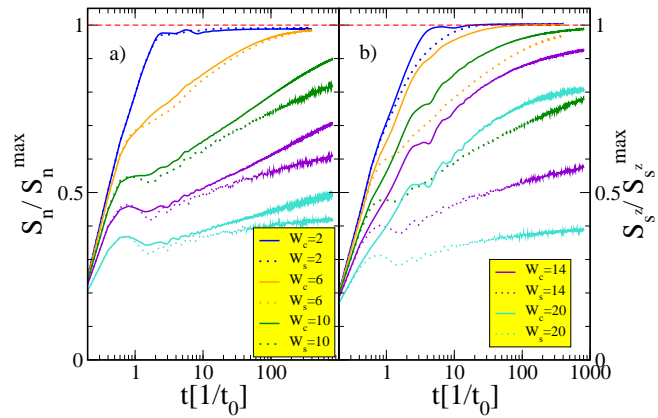}
\caption{$S_n(t)$ in a) and $S_{S^z}(t)$ in b) for different strengths of disorder. Full lines represent  potential disorder $W_c$ while dotted lines magnetic field disorder $W_s$. Results are normalized to their respective  values in the infinite--$T$ limit, $S_{n}^\mathrm{max}\sim 1.45$ and $S_{S^z}^\mathrm{max}\sim 1.80$.  Parameters of the system are the same as in Fig.~1.  }
\label{Fig4}
\end{figure}

The charge entropy $S_n$, shown in Fig.~\ref{Fig4}(a), displays  qualitatively similar behavior with respect to either the potential or  the  field  disorder for small $W_{c(s)}\leq 6$. In both cases $S_n$ approaches its maximal value, characteristic for a thermal state at  $t\to \infty$. For larger $W_{c(s)}\geq 10$  we observe a stronger deviation for different types of disorder in the long--time limit. In both case se observe  a slow  logarithmic increase, characteristic for  MBL systems.%

In contrast, the spin entropy $S_{S^z}$ quantitatively differs in comparison to potential  or field  disorder in the whole range of $W_{c(s)}$'s. The most significant difference is observed when comparing results for larger $W_{c(s)}\geq 10$ where $S_{S^z}(W_c)$ grows significantly faster than $S_{S^z}(W_s)$. For example: while at $W_{c(s)}=10$ $S_{S^z}(W_c)$ nearly reaches its maximal value $S_{S^z}^\mathrm{max}$, $S_{S^z}(W_s)$ displays slow logarithmic growth.  For even larger $W_{c}\geq 14$ there seems to be no observable time interval with a logarithmic growth of $S_{S^z}(W_c)$. In contrast, it shows a tendency towards saturation towards a non--thermal value.

  To gain a deeper physical picture  we first note that while the field  disorder affects charge as well as spin degrees of freedom, the potential  disorder affects only charge degrees of freedom. For example: when a fermion with spin--up hops between neighboring sites under the influence of field disorder, it feels different Zeeman energy just as when in the presence of the potential  disorder it feels different potential energy.  There exist connected spin chains separated by empty sites. Spins that form a particular connected spin chain do not feel the potential  disorder as long as they remain attached to the chain. Spin excitation  can freely propagate along a connected spin chain in the presence of the potential  disorder. When such connected spin chains extend between the boundaries of the two subsystems, they contribute to the growth of spin entropy. 

\section{Spectral level statistics}

\begin{figure}
\centering
\includegraphics[width=\columnwidth]{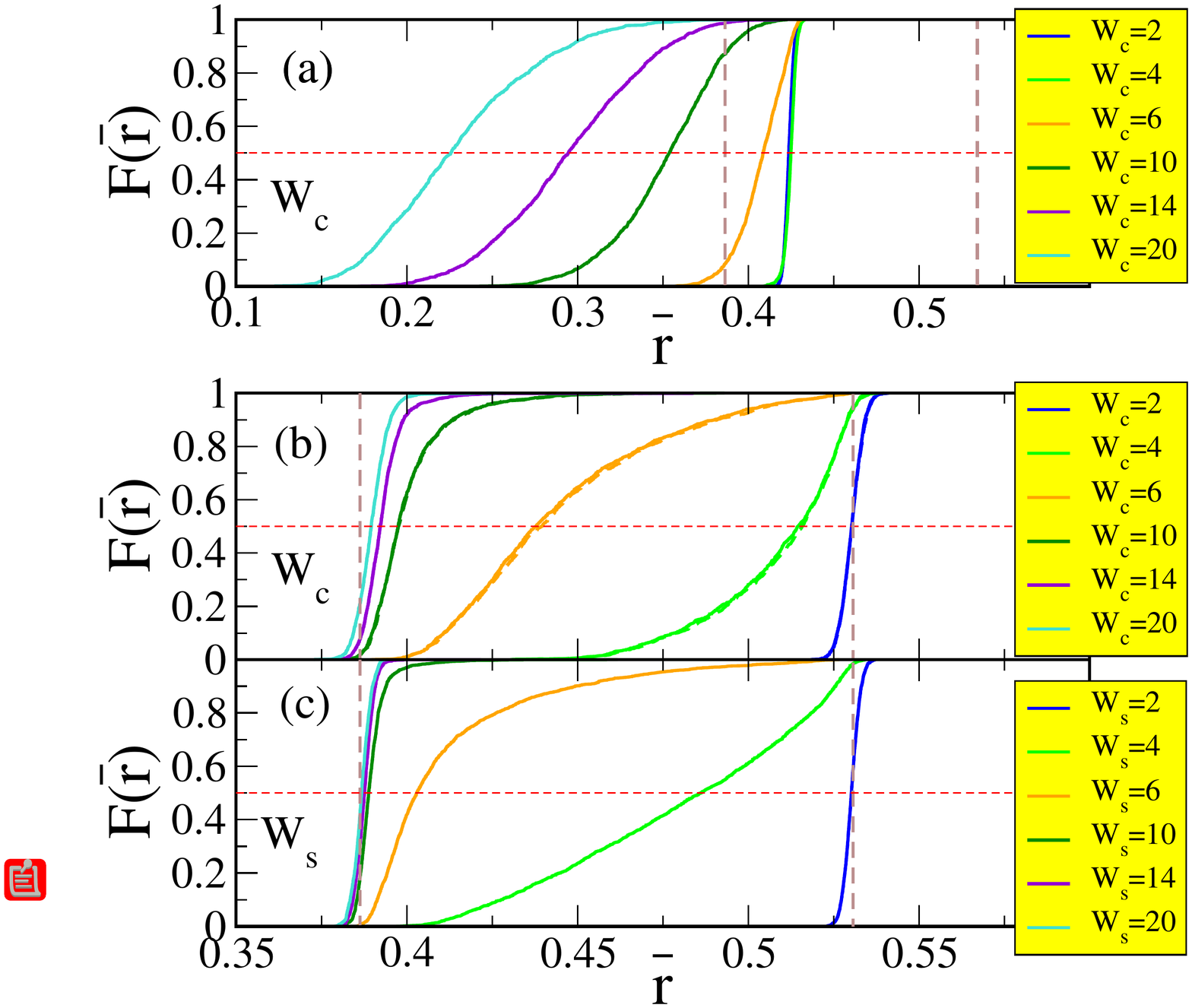}
\caption{ $F(\bar r)$ for different values of potential  disorder in (a) from the full spectrum, (b) from the symmetric (full lines) and the antisymmetric (dashed lines) part of the spectra and for field  disorder in (c). Vertical dashes lines indicate values of 
$r_\mathrm{Poisson} \simeq 0.386$ and $r_\mathrm{GOE} \simeq 0.53$. In most cases  $N_r\sim 1500$ samples have been used. Results were obtain using full diagonalization on a system with  $L=12$, $N_\uparrow=N_\downarrow = 3$ and $N_\mathrm{states}=18480$.}
\label{Fig5}
\end{figure}

Motivated by the rather unexpected difference in the time evolution of charge vs. spin imbalance, observed in Fig.~\ref{Fig1}(b) as well as other observables  probing  charge or spin degrees of freedom  under the influence of potential  disorder, we next explore statistical properties of the energy spectra. We compute adjacent energy level   spacing ratios, $r_n =\mathrm{min}[\Delta_n,\Delta_{n-1}]/\mathrm{max}[\Delta_n,\Delta_{n-1}]$ where $\Delta_n = E_n - E_{n-1}$ and $\{E_n\}$ represents the ordered set of energy levels of Hamiltonian in Eq.~(\ref{ham}).  For each realization of disorder we compute average value of $\bar r$ and then instead of computing average over different realizations, we calculate  the cumulative distribution function for $\bar r$, $F(\bar r)$. In Fig.~\ref{Fig5}(a) we first present $F(\bar r)$ for the case of potential disorder, $W_c$, taking into account the full spectrum. Since the Hamiltonian in Eq.~(\ref{ham}) is non--integrable, one expects that at small $W_c=2$ its spectrum resembles the spectrum of the Gaussian random matrices where $\bar r_\mathrm{ave} = \bar r_\mathrm{GOE} \simeq 0.53 $\cite{mukerjee_2006,oganesyan_huse_07}.  In contrast, at large $W_c = 20$, the average value of $\bar r$ should not drop below $r_\mathrm{Poisson}=0.386$, characteristic for a random distribution of energy levels. Distributions $F(\bar r)$, shown in Fig.~\ref{Fig5}(a) are not consistent with either  of the above predictions. 

For a proper analysis of the spectral level statistics in the case of the potential disorder we have computed $F(\bar r)$ separately for each symmetry subspace. In Figs.~\ref{Fig5}(b) and \ref{Fig5}(c)   we present $F(\bar r)$ for different values of $W_{c(s)}$. There are two types of nearly overlapping curves (full and dashed lines) in the case of potential disorder, see Fig.~\ref{Fig5}(b) representing separately $F(\bar r)$  for each symmetry sector. For small $W_{c(s)}=2$, presented in Figs.~\ref{Fig5}(b) and (c), $F(\bar r)$ can be fitted with the Error  function positioned  at $\bar r_\mathrm{ave}=0.53$,  which agrees with $r_\mathrm{GOE}$. At large $W_{c(s)}=20$ $F(\bar r)$ again resemble  Error functions, however in this regime  close to    $r_\mathrm{Poisson}$, which on a finite system indicate localization. For the intermediate  values of $W_{c(s)}\in [4,6]$ we observe broad distributions $F(\bar r)$ that  result from a mixture of systems where a part of them exhibit level statistics that resembles ergodic systems  while others the one closer to  non--ergodic/localized. 

We have also calculated the  distribution of the gap ratios without averaging $r_n$ over multiple energy levels. To this end we have generated a set containing $r_n$ for various
 $n$ as well as for various disorder realizations and calculated the probability density, $P(r)$, from the histogram of $\{ r_n\}$. While the distribution $F(\bar{r})$ in Fig. \ref{Fig5} contains information about differences between various  realizations
of disorder, such information is not directly encoded in $P(r)$. Nevertheless,  $P(r)$ allows for a more detailed comparison with the random matrix theory. 
 For the  Poisson level statistics one gets \cite{oganesyan_huse_07} 
\begin{equation}
P_{P}(r)=\frac{2}{(1+r)^2},
\label{pp}
\end{equation} 
 while an approximate  distribution for the Gaussian-Orthogonal- Ensemble (GOE) can be derived from the  Wigner surmise \cite{atas2013,alet2022,fremling2022} for 3 energy levels
 \begin{equation}
P_{GOE}(r)=\frac {27}{4} \frac{r+ r^2}{(1 + r + r^2)^{5/2}}.
\label{pgoe}
\end{equation} 

\begin{figure}
\centering
\includegraphics[width=\columnwidth]{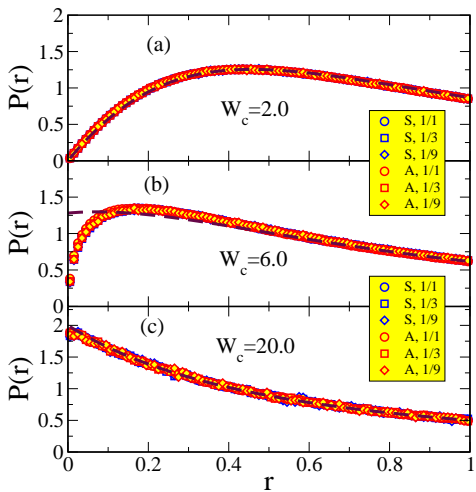}
\caption{Points show the probability density of the gap ratio  for charge disorder obtained from symmetric (S) or antisymmetric (A) parts of spectra.  
The fractions in the legend mark the parts of spectra which were used for generating the distribution. The dashed  lines in (a) and (c) show Eqs. (\ref{pgoe}) and (\ref{pp}), respectively.
The dashed line in (b) shows distribution given by Eq. (20)  in Ref. \cite{fremling2022} with a normalization that is relevant for $r \in (0,1)$. }
\label{Fig6}
\end{figure} 
\begin{figure}
\centering
\includegraphics[width=\columnwidth]{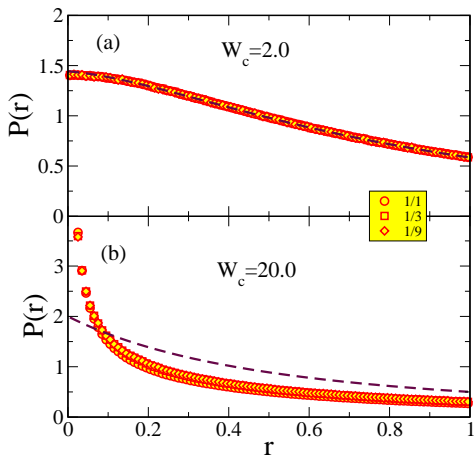}
\caption{  The same as in Fig. \ref{Fig6} but for a full spectra containing symmetric and antisymmetric levels.
The dashed line in (a) shows distribution given by Eq. (24)  in Ref. \cite{fremling2022} with a normalization that is relevant for $r \in (0,1)$, whereas the dashed line in (b) shows Eq. (\ref{pp}).
 }
\label{Fig7}
\end{figure} 

Figure \ref{Fig6} shows the distributions, $P(r)$, obtained from  the symmetric (S) or the antisymmetric (A) parts of spectra with charge disorder.  
In order to identify artifacts arising from the presence of the localization edge, 
the distributions have been  generated either from  all levels or only from a central part ($1/3$ or $1/9$) of the levels in the middle of the spectrum. Results obtained for all three cases accurately overlap 
(see Figs.  \ref{Fig6}  and  \ref{Fig7} ) indicating absence of artifacts originating from the localization edge.

As expected, numerical results for weak disorder [Fig. \ref{Fig6}(a)]  accurately reproduce  Eq. (\ref{pgoe})
whereas for strong disorder shown in  Fig. \ref{Fig6}(c),  the distribution agrees with Eq. (\ref{pp}). In the vicinity of the transition, $P(r)$ can be well approximated by a mixed Wigner surmise discussed  very recently in Ref. \cite{fremling2022}. 
More precisely, the dashed curve in  Fig. \ref{Fig6}(b) shows  distribution for $2 \times 2$ GOE matrices mixed with two uncorrelated energy levels, see Eq. (20)  in Ref. \cite{fremling2022}.
Such mixture of GOE and Poisson distributions may be interpreted as a coexistence  of  localized (insulating) and delocalized (metalic) domains whereby the absence of level crossings, $F(r \to 0) =0$ in Fig. \ref{Fig6}(b), indicates that localization within the former domains is not perfect. Similar results concerning the spatial separation of conducting and insulating domains have been recently found
 for the random-field Heisenberg model \cite{herbrych2022}.

Figure \ref{Fig7} shows the distribution of the gap ratios obtain for charge disorder from the full spectrum that includes both  symmetric and antisymmetric levels. Results obtained for weak disorder can be very accurately approximated  by two independent GOE ensembles. In particular,  dashed line in  Fig. \ref{Fig7}(a) shows the distribution derived for mixture of two  $2 \times 2$ GOE matrices, see Eq. (24)  in Ref. \cite{fremling2022}. Rather unexpected is the case of strong disorder when  $P(r)$  for $r \ll 1$ significantly exceeds the distribution derived for the Poisson statics, as it is shown in Fig. \ref{Fig7}(b).
Comparing Fig. \ref{Fig6}(c) and  \ref{Fig7}(b) one identifies an attraction between  symmetric and antisymmetric energy levels.
Such scenario is consistent with results for the average ratio shown in Fig. \ref{Fig5}(a).

\section{ Summary}
We have studied spin- and charge-dynamics in a disordered chain such that an unperturbed system has $\mathbb{Z}_2$ spin-symmetry. Then we considered two types of disorder: a random magnetic field that breaks the $\mathbb{Z}_2$ symmetry and
random charge potential, which preserves the spin-symmetry. In the former case with broken spin-symmetry, the dynamics of all studied observable are consistent with localization on finite lattices in that their expectation values do not approach the results for thermal equilibrium.
 However, for the symmetry-preserving disorder, the 
observables that are odd under the $\mathbb{Z}_2$ spin transformation seem to thermalize while even observables do not. Numerical studies of the time-evolution for the symmetry-preserving disorder have been accompanied by the level statistics. Interestingly, the level statistics 
obtained separately for odd and even symmetry sectors accurately reproduce a crossover/transition from the GOE for weak charge disorder to the Poisson distribution for the strong disorder. The time evolution and the level statistics suggest that localization exists within each symmetry sector, i.e. for odd or even eigenstates and observables with matrix elements only within a given symmetry sector. The apparent relaxation of odd operators is not inconsistent with the level statistics since such operators evolve between the two sectors. Similar numerical results have been found for the dynamics in the Hubbard model with charge disorder, which, however, has SU(2) symmetry\cite{prelovsek16,kozarzewski18,sroda19,kozarzewski2019,protopopov2019}. In the latter model, the spin imbalance decays subdiffusively \cite{prelovsek16,kozarzewski18,sroda19} while the spin-energy-density seems not to thermalize \cite{kozarzewski2019,protopopov2019}. 
In our studies, we have not considered the stability of the localized phase in the thermodynamic limit, so we do not exclude that localization represents extremely slow dynamics that eventually may lead to a thermal equilibrium of an infinite chain.

 Comparison of the entanglement entropies of spin and charge degrees of freedom reveals an essential difference between the field disorder that affects charge and spin degrees of freedom and the potential disorder that influences only charge degrees of freedom.  Spins that form a particular connected spin chain do not feel potential disorder. Spin excitations can thus freely propagate along a connected spin chain. This may explain the absence of a logarithmic growth of the spin entanglement entropy even in the regime of strong potential disorder, where in contrast, the charge entanglement entropy displays logarithmic time evolution.

The original motivation for this work stems from experiments on cold atoms \cite{schreiber_hodgman_15} where charge imbalance was measured in cold atoms experiment setup. Recent advances in the spin- and density--resolved microscopy\cite{Bloch2016,Bloch2020,Bloch2020a} might allow measurements of the charge and the spin imbalance under the non--symmetry--breaking disorder.

\acknowledgements
We acknowledge the support by the National Science Centre, Poland via project 2020/37/B/ST3/00020 (M.M.), the support by the Slovenian Research Agency (ARRS), Research Core Fundings Grants P1-0044 ( J.B.) and the support from  the Center for Integrated Nanotechnologies, an Office of Science User Facility operated for the U.S. Department of Energy (DOE) Office of Science by Los Alamos National Laboratory (Contract 89233218CNA000001) and Sandia National Laboratories (Contract DE-NA-0003525) (J.B.). 
\bibliographystyle{biblev1}
\bibliography{references,references_ergtransition,Janez_bib}

\end{document}